\documentclass[twoside]{ilcws08}
\usepackage[latin1]{inputenc}
\usepackage[pdftex]{graphicx,epsfig,color}
\usepackage{wrapfig,rotating}
\usepackage{amssymb,amsmath,array}
\usepackage{subfigure}

\begin{document}
\title{
A TPC with Triple-GEM Gas Amplification and TimePix Readout} 
\author{
Uwe Renz\thanks{University of Freiburg - Department of Physics, Hermann-Herder-Str. 3, 79104 Freiburg - Germany,  renz@physik.uni-freiburg.de}\\ \textbf{-On behalf of the LCTPC Collaboration-}
}

%

%


\maketitle

\begin{abstract}
We present measurements with a GEM based TPC prototypes and the TimePix CMOS ASIC as charge collecting anode. Tests with a small prototype  were performed in a 5\,GeV electron beam at DESY. 
It is shown that the point resolution for short drift distances is better than 20\,$\mu \mathrm{m}$. The achieved time resolution is better than 10\,ns at 100\,MHz clock frequency. Furthermore, experimental studies with an enlarged pixel size are addressed.
To study the performance for longer drift distances a chamber with 26\,cm drift length is operated in Bonn. Using also GEM and TimePix for the readout, data with cosmic muons and a Sr$^{90}$ source are recorded. This chamber uses also GEMs and TimePix for the readout. data with cosmic muons and a Sr90 source are recorded. The dependency of spatial resolution and cluster properties on the drift distance are studied.

\end{abstract}

\section{Introduction}
The development of Micro-Pattern-Gas-Detectors (MPGDs) expands the field of application for detectors based on gas multiplication. 
In particular, a large Time Projection Chamber (TPC) wit MPGD readout is foreseen in the ILD detector concept for the ILC \cite{ILCILD}. R\&D towards such a TPC is carried out within the LCTPC collaboration \cite{LCTPC_coll}
For the prototypes presented here the conventional pad readout of GEMs, with a typical pad size in the order of a few $\mathrm{mm^2}$, is replaced by a highly integrated device, the TimePix CMOS pixel ASIC \cite{LlopartCudie:2007zz}. The advantage of this concept is the small pixel size of the TimePix, with $ 55\,\mathrm{\mu m} \times 55\,\mathrm{\mu m}$ matching the typical granularity of the GEMs.
Tests with the triple GEM/TimePix setup developed at the University of Freiburg have been performed in a 5\,GeV electron beam at DESY. The Bonn setup consists of a TPC fieldcage with 26\,cm drift length. This allows to study cluster and track properties for longer drift distances.

\section{Setups}
\subsection{GEMs and TimePix}
\subsubsection{GEMs}
\begin{figure}[!t]
\centerline{
\subfigure[]{\includegraphics[width=4.5cm]{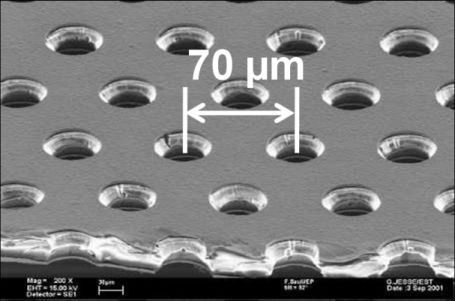}
\label{GEM_RTM}}
\hfil
\subfigure[]{\includegraphics[width=4.5cm]{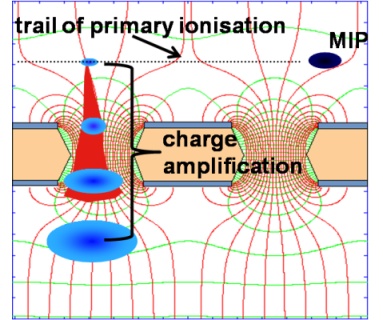}
\label{GEM_Garf}}
\subfigure[]{\includegraphics[width=3.75cm]{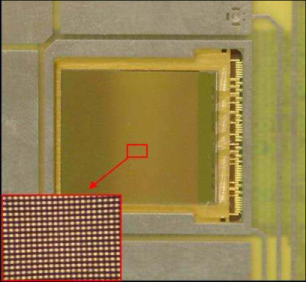}
\label{tpx_inlet}}
\hfil
}
\caption{a) Scanning tunneling microscope image of a standard GEM.\newline b) A simulation of electrostatic potential (green) and field (red) for two GEM holes \cite{GEM_ref}. Above the GEM a Minimum Ionizing Particle (MIP) is represented by a black dot crossing from the left to the right, leaving a trail of primary ionization. \newline c) Image of a TimePix chip installed in the setup. The inlet shows a magnified view of the chip, where single pixels are visible.}
\label{GEM_Fig}
\end{figure}

A GEM \cite{GEM_ref} is a thin metal-coated polymer foil with a high density of chemically etched holes arranged in a regular pattern. Two different pitch-geometries are investigated. First standard GEMs, with 70\,$\mathrm{\mu m}$ \ hole diameter and 140\,$\mathrm{\mu m}$ \ hole pitch had been investigated. Then tests with a smaller pitched GEM \cite{Bellazzini:2006bg} were performed. These GEMs, like standard GEMs, have the same thicknesses for the Kapton$^\text{\textregistered}$ and copper layers. However the holes have an outer diameter of 30\,$\mathrm{\mu m}$ \ and a pitch of 50\,$\mathrm{\mu m}$ \ and the active surface is $28\,\mathrm{mm} \times 24\,\mathrm{mm}$. Fig. \ref{GEM_Garf} shows the electrostatic potential (green) and field (red) for two GEM holes. 

\subsubsection{TimePix}
The TimePix CMOS readout pixel chip is used as highly segmented charge collecting anode. Fig. \ref{tpx_inlet} shows a TimePix chip installed in the setup. On the chip 65536 pixels are arranged in a square matrix of $256 \times 256$ lines and columns. As the size of each pixel cell is $55\,\mathrm{\mu m} \times 55\,\mathrm{\mu m}$ the total active surface is $14\,\mathrm{mm} \times 14\,\mathrm{mm}$. A reference frequency is distributed throughout the entire chip.
For each pixel one of the following four modes can be set individually:
\begin{description}
	\item[TOT:] In the ``Time-Over-Threshold'' mode measures time a signal is above an adjustable threshold.
	\item[TIME:] In the TIME-mode measures time from the point the signal crosses the adjusted threshold until a common stop by a gate signal called ``shutter''.
	\item[MediPix:] \ \ \ In the MediPix-mode the number of hits crossing an adjustable threshold are recorded.
	\item[OneHit:] \ \ The first signal crossing the adjusted threshold sets the counter to one. Any further hits will be ignored.
\end{description}
To record simultaneously TIME and TOT information the TimePix is operated in a special mode: In a checkerboard-like fashion consecutive pixels are set to TOT and TIME mode. Consequently half of the pixels record TIME, while the other half records TOT. 

\subsection{Freiburg Test Beam Setup}
\begin{figure}[!t]
\centerline{
\subfigure[]{\includegraphics[width=7.0cm]{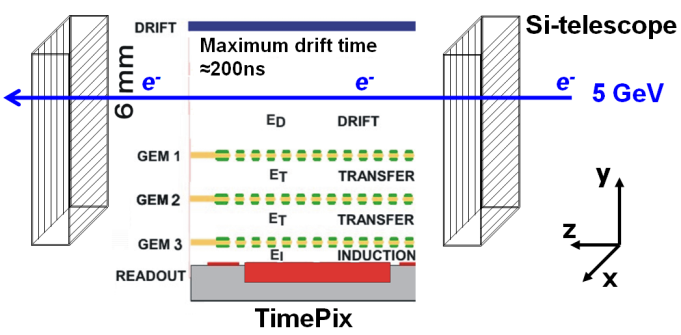}\label{stack_tel_Fig}}
\subfigure[]{\includegraphics[width=3.50cm]{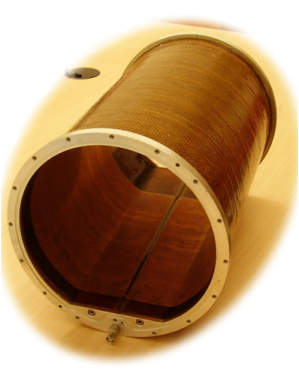}\label{BN_fieldC}}
}
\caption{a) Shown is a vertical cut of the triple GEM stack with the TimePix chip and the two Si-strip telescope layers used for an external
 determination of tracks.\newline b) TPC field cage.}

\end{figure}

The TimePix is positioned at the underneath a triple GEM stack as shown in Fig. \ref{stack_tel_Fig}. 
The GEM stack is operated at a nominal gain of $\approx 10^5$ with an $\mathrm{Ar/CO_2}$ 70/30 gas mixture.  

The 5\,GeV electron beam transverses a volume with 6\,mm drift distance parallel to the GEM planes. 
As displayed in Fig. \ref{stack_tel_Fig} the setup is positioned in between the two sensor layers of a Si-strip-telescope \cite{Milite:2001wr}. The telescope has a 50\,$\mathrm{\mu m}$ \ readout pitch and is used for external track determination.


\subsection{Bonn TPC laboratory}
The Bonn setup consists of a field cage with a diameter of 26\,cm and a maximum drift length of 26\,cm. This is shown in Fig. \ref{BN_fieldC}. The maximum drift field, which can be applied is 1\,kV/cm. The total material budget of the field cage is about 1\%\,$X_0$. The coincidence of two scintillator counters is used as trigger for cosmics. A $\mathrm{Sr^{90}}$-source is also used for specific measurements. The readout is performed using a similar scheme as in the Freiburg setup. 
\begin{figure}[!t]
\centering
\includegraphics[width=8.5cm]{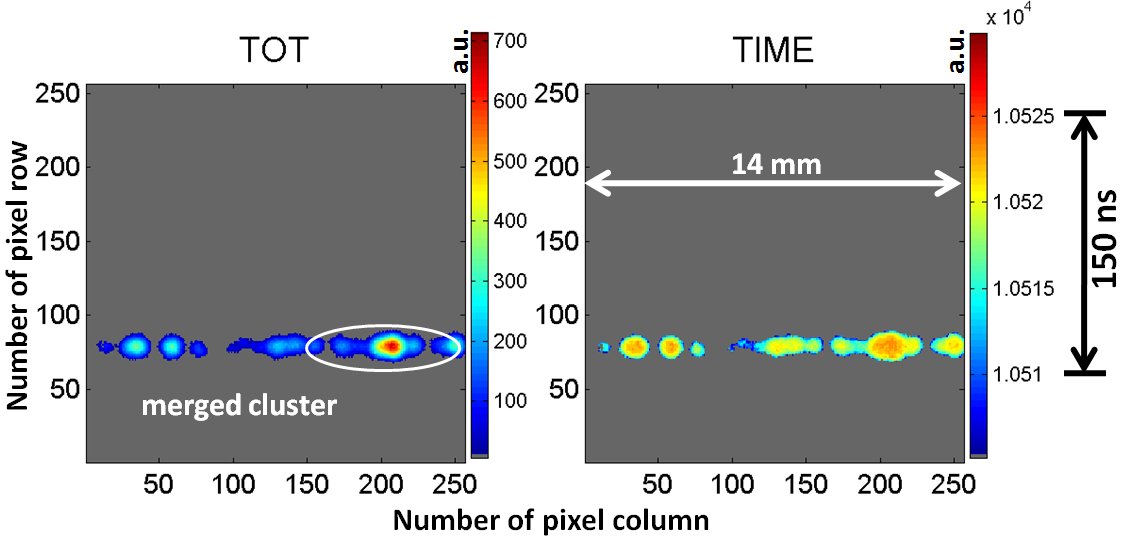}\label{eventDispay}
\caption{Image of a typical event. Left in TOT mode and right in TIME mode.}
\end{figure}

\section{Results}
\subsection{Typical Events and cluster reconstruction}
In Fig. \ref{eventDispay} a typical event is shown. Using an adapted ZEUS cluster splitting algorithm \cite{ZEUS} typically $\approx 11$ clusters are reconstructed in an event, while from simulations \cite{Hauschild} about 40 clusters are expected for Ar/CO$_2$ 70/30.


The evaluation followed the methods presented in \cite{Bamberger:2006xp}: First a straight line was fitted to all $N$ cluster centroids and the respective N residuals were calculated. After this one cluster had been exempted and a fit to the remaining $N-1$ clusters was performed. The deviation was calculated for the excluded cluster and then the method was repeated for all remaining permutations of clusters. 
From the width of both residual-distribution, denoted $\sigma_N$ and $\sigma_{N-1}$, the geometric mean $\sigma_{mean} = \sqrt{\sigma_{N} \times \sigma_{N-1}}$ was calculated. This yields an unbiased estimator for the resolution.

\subsection{Analysis of Test Beam data}
\subsubsection{Spatial Resolution}
\begin{figure}[!t]
\centerline{
\subfigure[]{\includegraphics[width=5.750cm]{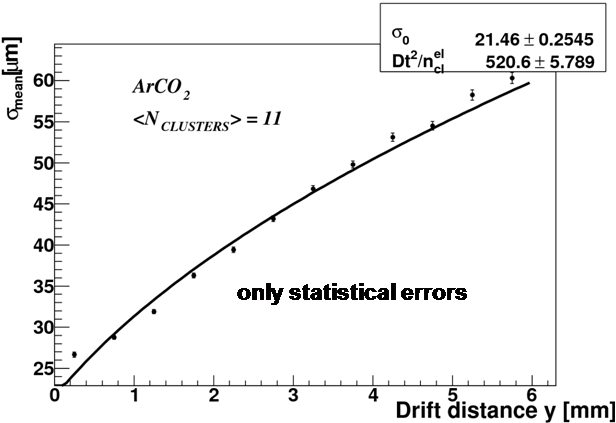}\label{resStdGems}}
\subfigure[]{\includegraphics[width=5.75cm]{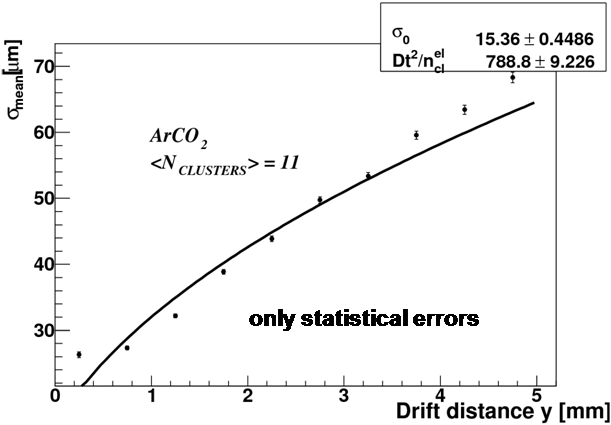}\label{resSmallGems}}
}

\caption{a) Resolution $\sigma_{mean}(y)$ as function of drift distance $y$ for standard GEMs.
b) Resolution $\sigma_{mean}(y)$ as function of drift distance $y$ for small pitched GEMs.}
\label{sigmay}
\end{figure}
The spatial resolution $\sigma_{trans}(y)$ depends on the transverse diffusion along the drift path: $\sigma_{trans}(y)=\sqrt{\sigma_0^2 + D_t^2\times y/n^{el}_{cl}}$\label{old_formula}, with $\sigma_0$ denotes the best achievable resolution for zero drift distance, $D_t$ the constant of transverse diffusion and $n^{el}_{cl}$ the number of primary electrons contributing to a cluster.\\
Using the information from the Si-strip telescope the events were sorted into bins of $0.5\,\mathrm{mm}$ along the drift distance $y$. The resolution at zero drift was extrapolated from a fit of $\sigma_{trans}(y)$ to the measured values $\sigma_{mean}(y)$ in Fig.\ref{sigmay}. Including the systematic uncertainty for $\sigma_0$ of about $\pm 1\,\mathrm{\mu m}$ due to the position of the Si-telescope relative to the GEM/TimePix detector, the result for the standard GEMs was $\sigma_0 = 22 \pm 2 \,\mathrm{\mu m}$ and for the smaller pitched GEMs $\sigma_0 = 15 \pm 1 \,\mathrm{\mu m}$.

\subsubsection{Time Walk Correction}
\begin{figure}[!t]
\centerline{
\subfigure[]{\includegraphics[width=5.0cm]{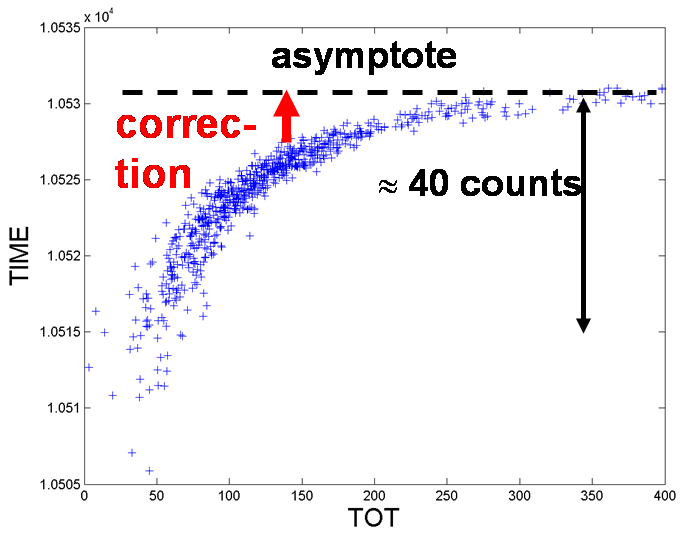}
\label{timeTotCorr}}
\subfigure[]{\includegraphics[width=5.0cm]{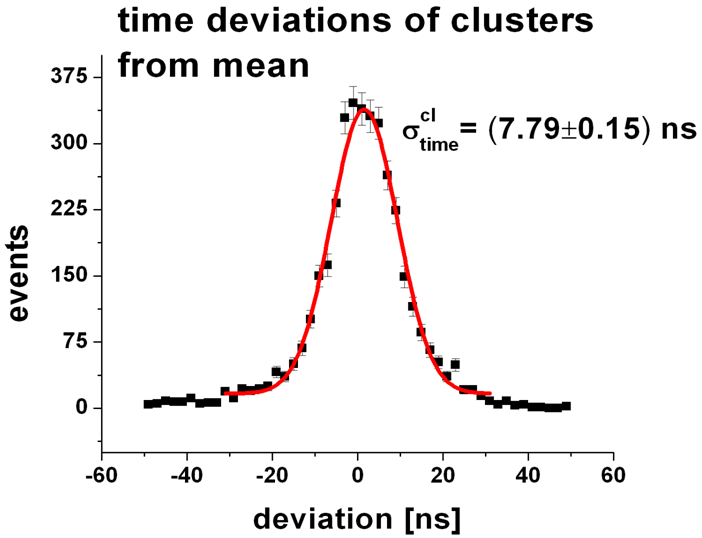}
\label{timeRes}}
}
\caption{a) Correlation of TOT vs. TIME. b) Plotted are the deviations of corrected TIME values from the corrected mean time in the corresponding event. A fit to these values results in a $\sigma_{time}$ of $\approx 8\,\mathrm{ns}$.}
\label{sigmatime}
\end{figure}

The TIME-measurements showed a dependency on the pulse height (time walk) as displayed in Fig. \ref{timeTotCorr}. TIME-values measured next to pixels with small TOT seemed to arrive systematically later compared to TIME-values which lay close to pixels with a large TOT-value.
From the TOT-TIME-correlation a correction function had been derived and was applied to the mixed mode data. Fig. \ref{timeRes} shows the deviations of (corrected) TIME-values at the center of a cluster relative to the mean time\footnote{This was determined per event by averaging over all TIME values at the cluster centroids.} of the corresponding event. The achieved time resolution is $8\,\mathrm{ns}$, which is smaller than the used clock period of 10\,ns. 

\subsection{Analysis of TPC-Lab Data}
\subsubsection{Resolution studies}
\begin{figure}[!t]
\centerline{
\subfigure[]{\includegraphics[width=6.0cm]{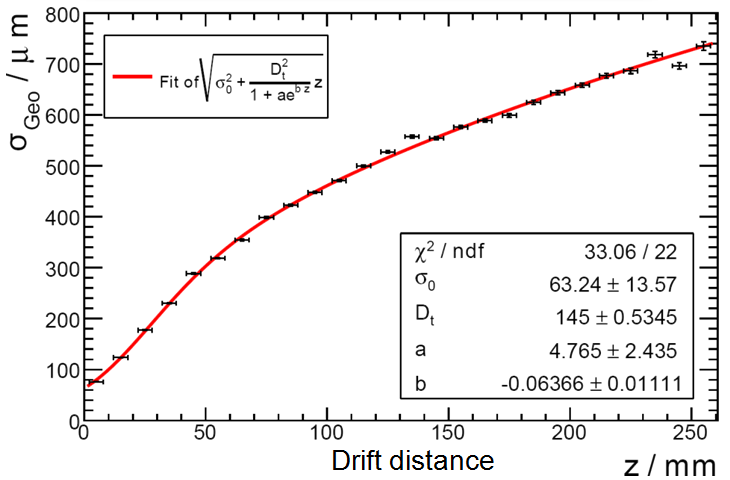}
\label{resoltution_for_long_drift}}
\subfigure[]{\includegraphics[width=6.5cm]{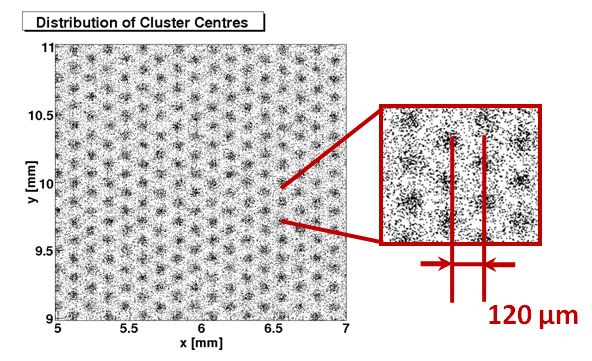}
\label{hole_pattern}}
}
\caption{a) Resolution as function of drift distance up to 26\,cm. c) Distribution of reconstructed cluster centroids revealing the discreteness of the GEM hole pattern.
}
\label{TIME_corr}
\end{figure}

With the Bonn setup, cluster and track properties could be studied over a much longer drift path. Analysis of this data \cite{Simonchen} revealed that the parametrization of $\sigma_{trans}(z)$ used for the test beam data does not describe the data for drift distances\footnote{In the Bonn setup the drift direction is denoted the $z$-axis.} above several centimeters. This suggests that the number of primary electrons contributing to a cluster is not fixed, as in \ref{old_formula}, but a function of $z$. For short drift distances, where the transverse diffusion is small, multiple electrons contribute to a cluster. In contrast to large drift distances, where the transverse diffusion is large and (each) individual electron forms its own cluster at the readout plane. Accordingly the parametrization for $n_{cl}^{el}(z)$ is described by:
\[
n_{cl}^{el}(z) = 1 + a \cdot e^{bz} \Rightarrow 
\left\{ 
\begin{array}[c]{l}
																													\lim\limits_{z \rightarrow 0}{n_{cl}^{el}(z)=a=5\pm 2,\text{taken from Fit in Fig.\ref{resoltution_for_long_drift}}}\\
																													\lim\limits_{z \rightarrow \infty}{n_{cl}^{el}(z)=1}, b<0
\end{array}
\right.
\]
A fit of $\sigma_{trans}(z)=\sqrt{\sigma_0^2+D_t^2/n_{cl}^{el}(z)}$ to the measured values $\sigma_{mean}$ from measurements with cosmics is shown in Fig. \ref{resoltution_for_long_drift}. The average resolution between $ 0\,\mathrm{mm} < z < 10\,\mathrm{mm}$ of $\sigma_{mean}(5\,\mathrm{mm})\approx 75\,\mathrm{\mu m}$ is in agreement with the results from the Freiburg setup with very short drift volume.\\ But with such a good resolution the discrete nature of the GEM holes and the readout starts to matter. This can be seen nicely from Figure \ref{hole_pattern}, where a two dimensional histogram of the reconstructed cluster centroids from a Sr$^{90}$  is shown. This reproduces the GEM hole structure and explains the improved $\sigma_0$ for the small pitched GEMs in Fig. \ref{resSmallGems}.

\subsubsection{Declustering}
\begin{figure}[!t]
\centering
\includegraphics[width=5.0cm]{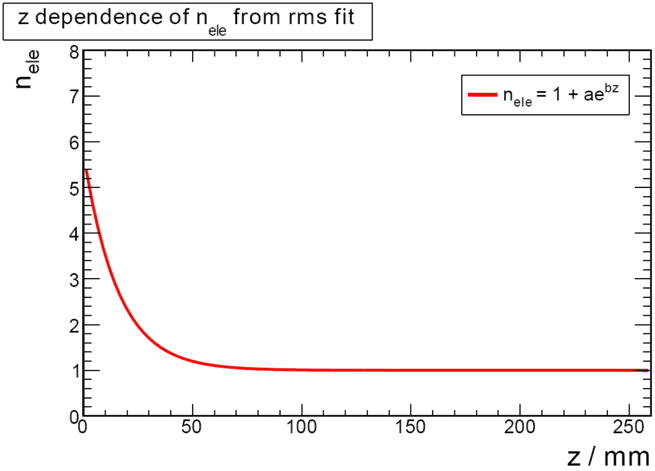}
\caption{
Trend of $n_{cl}^{el}(z)$ as function of $z$.}\label{n_eff_vs.z}
\end{figure}

\begin{figure}[!t]
\centerline{
\subfigure[]{\includegraphics[width=5.00cm]{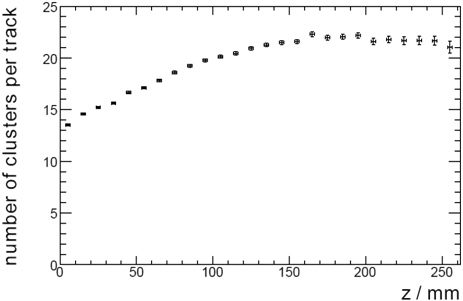}
\label{number_of_cluster}}
\subfigure[]{\includegraphics[width=5.250cm]{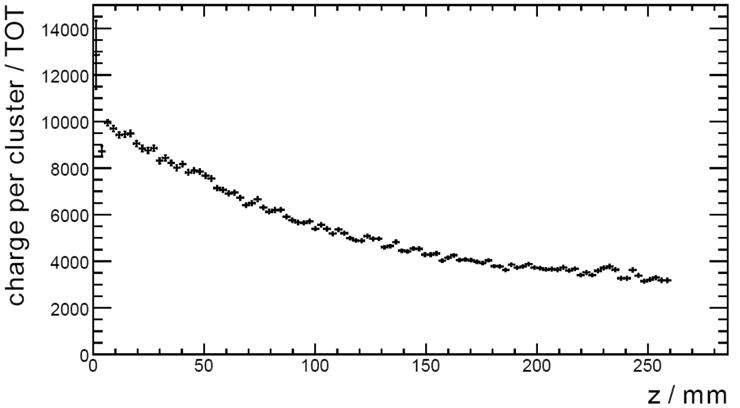}
\label{charge_per_cluster}}
}
\caption{
a) Number of reconstructed clusters per track versus drift distance.
b) Charge per reconstructed cluster versus drift distance.}
\label{BN_declustering}
\end{figure}

The hypothesis of a decreasing $n_{cl}^{el}(z)$, shown in Fig. \ref{n_eff_vs.z}, is supported by the plots in Figure \ref{BN_declustering}. Fig. \ref{number_of_cluster} shows how the number of reconstructed clusters per track increases for larger drift distances. At the same time the charge per cluster in Fig. \ref{charge_per_cluster} decreases for larger $z$. As indicated before, this observation can be explained with two arguments: First, for short drift distances, several primary electrons contribute to a reconstructed cluster. Hence the number of distinct clusters is small, but the charge per cluster is large. Second, for drift distances above several centimeters, the transverse diffusion is large enough to separate primary electrons far enough, so that they are reconstructed as individual clusters. Consequently the charge per cluster takes its minimum, while the number of clusters is maximal (``declustering'').

\section{Conclusion}
In a DESY test beam it was shown that the point resolution for short drift distances is better than 20\,$\mathrm{\mu m}$. The achieved time resolution is better than 10\,ns at 100\,MHz clock frequency. Measurements with a larger TPC prototype in the Bonn TPC laboratory were performed for drift distances up to 26\,cm. A detailed analysis of these data revealed the declustering effect for long drift distances beyond several centimeters.

\section*{Acknowledgements}
We would like to thank the TPC group of the RWTH Aachen for the design and the construction of the field
cage, our colleagues from NIKHEF, Amsterdam, the University of Prague and the Freiburg Material Research Center for supporting the 
commissioning of the readout electronics.\\
This work is partly supported by the Commission of the European Communities under the 6th framework program 
"Structuring the European Research Area", contract number RII3-026126 and by the German Federal Ministry 
of Education and Research under Grant No. HS6PD2.


\begin{footnotesize}



%

\end{footnotesize}


\end{document}